\documentclass[acmtog]{acmart}
\renewcommand\footnotetextcopyrightpermission[1]{} 
\AtBeginDocument{%
  \providecommand\BibTeX{{%
    \normalfont B\kern-0.5em{\scshape i\kern-0.25em b}\kern-0.8em\TeX}}}

\setcopyright{acmcopyright}
\copyrightyear{2019}

\usepackage{amsmath} 
\usepackage{amssymb}  
\usepackage{algorithm2e}
\usepackage{graphicx}
\usepackage{subfig}
\usepackage{float}

\begin{document}

\title{Semantic Image Completion and Enhancement using Deep Learning}

\author{Vaishnav Chandak}
\email{vaishnavchandak1996@gmail.com}
\affiliation{%
  \institution{ABV-Indian Institute of Information Technology and Management, Gwalior}
  \city{Gwalior}
  \state{Madhya Pradesh}
  \country{IN}
  \postcode{474015}
}

\author{Priyansh Saxena}
\affiliation{%
  \institution{ABV-Indian Institute of Information Technology and Management, Gwalior}
 \city{Gwalior}
  \state{Madhya Pradesh}
  \country{IN}}
\email{saxenapriyanshasd@gmail.com}
\orcid{0000-0003-1407-9752}

\author{Manisha Pattanaik}
\affiliation{%
  \institution{ABV-Indian Institute of Information Technology and Management, Gwalior}
\city{Gwalior}
  \state{Madhya Pradesh}
  \country{IN}}
\email{manishapattanaik@iiitm.ac.in}

\author{Gaurav Kaushal}
\affiliation{%
  \institution{ABV-Indian Institute of Information Technology and Management, Gwalior}
\city{Gwalior}
  \state{Madhya Pradesh}
  \country{IN}}
\email{kaushalg@iiitm.ac.in}

\renewcommand{\shortauthors}{Vaishnav Chandak, Priyansh Saxena, Manisha Pattanaik and Gaurav Kaushal}

\begin{abstract}
  In real life applications, certain images utilized are corrupted in which the image pixels are damaged or missing, which increases the complexity of computer vision tasks. In this paper, a deep learning architecture is proposed to deal with image completion and enhancement. Generative Adversarial Networks (GAN), has been turned out to be helpful in picture completion tasks. Therefore, in GANs, Wasserstein GAN architecture is used for image completion which creates the coarse patches to filling the missing region in the distorted picture, and the enhancement network will additionally refine the resultant pictures utilizing residual learning procedures and hence give better complete pictures for computer vision applications. Experimental outcomes show that the proposed approach improves the Peak Signal to Noise ratio and Structural Similarity Index values by 2.45\% and 4\% respectively, when compared to the recently reported data.
\end{abstract}



\maketitle

\section{Introduction}
Image completion and enhancement are two prime issues in the field of image processing and machine learning. In the past couple of years, the incredible developments of deep learning on different issues in low level as well as in high-level computer visions applications have been seen.  The low level computer vision issues includes picture completion \& its enhancement and are much foreseen to happen during image processing. Numerous image processing based techniques are proposed to take care of low-level vision issues. Many of them handle these issues separately \cite{dummy:3}, \cite{dummy:6}; however, most of the time image completion and denoising issues occur simultaneously.  
Many real world computer vision tasks endure from missing and masked areas in images which leads to poor quality images and hence the complexity of corresponding computer vision tasks increases. These problems are difficult to deal as estimating the missing region in the image is not easy \cite{dummy:1}.\\

Image completion refers to filling missing regions in the image based on the available visual data. On the other hand, image enhancement attempts to eliminate unwanted noise and blur from the image along with sustaining most of the image details.
In this paper, an efficient image completion and enhancement model is proposed, which intends to recover the corrupted and masked regions in images and then refining the image further to increase the quality of output image. The method is motivated by generative adversarial networks (GANs) \cite{dummy:8},  \cite{dummy:11}. Wasserstein GAN architecture of the GANs is used for image completion which recovers the missing regions by filling the corrupted part in the damaged image, and the enhancement network will further refine the completed images using residual learning techniques and then provide a better quality image as output.

\section{RELATED WORK}
\noindent Yizhen Chen and Haifeng Hu have proposed an upgraded approach for semantic inpainting of images. The proposed method named progressive inpainting using generative models in which they first estimated corrupted image distribution and then moderately refined image details. However, the model still could not handle large missing regions in images \cite{dummy:3}. Jia-Bin Huang and Ahuja have proposed an advance-knowledge approach which used contextual information for image completion. However, in case the corrupted region is large or is irrelevant to visual data, or if the complexity of the image is high, the output of the method would be quite unsatisfactory \cite{dummy:6}. Connelly Barnes and Eli have proposed patch matching algorithm for image completion for nonparametric texture construction. The algorithm performed satisfactorily and was able to identify similar patches. However, it failed when the original image lacked adequate data to complete the missing regions \cite{dummy:1}. Yunjin Chen and Thomas Pock have proposed nonlinear response dispersion model, which consists of a feed forward network with a fixed number of gradient descent stages. Trainable nonlinear reaction-diffusion accomplished promising execution in any case; its display was prepared for a specific noise level. It was unfit to perform well on pictures with obscure noise levels. Additionally, it requires the output which is expected by the network during training \cite{dummy:2}. In 2011, Deng transformed the inpainting task to the graph-labeling task using graph Laplace method. However, this method required images samples of the image to be inpainted be included in the training data, which was not practical in real life applications \cite{dummy:4}. A viable face inpainting algorithm utilizing a generative model was proposed by Yijun Li, Sifei Liu, and Jimei Yang. From background inpainting task, face inpainting is a challenging task because it regularly needs to produce semantically newer pixels areas in the missing region parts like eyes and nose, which can vary from person to person. Even though the model had the capacity to produce semantically conceivable and outwardly satisfying content, it has a few constraints. The model still could not deal with some unaligned faces. also, it did not wholly misuse the spatial connections between nearby pixels \cite{dummy:8}. Ruijun and Yang proposed an improved generative translation model. The paper proposed a semantic image completion method using regional completions for painting completion. Using the generator and discriminator network, the missing region is generated, which should be consistent with the surrounding region. However, image completion work is restricted to only face data and needed to be improved to ensure that the entire painting work could be recovered \cite{dummy:9}. Deepak Pathak put forward Context Encoders(CE) which estimated missing areas in images based on its surroundings. However, during training it needed a mask on the corrupted regions of the image, that is a significant disadvantage of the approach, and also context encoders led to blurry and noisy results in the inpainted parts \cite{dummy:12}.  Peyr proposed an adjustable low-dimensional manifold for images. It included inpainting task on synthetic as well as texture data. However, the employed work was quite away in giving solutions to real-world images of a face \cite{dummy:13}. Ren proposed a novel CNN architecture named Shepard Convolutional Neural Networks which efficiently equips conventional CNN with the ability to learn missing data. However, in case the corrupted region was large or was irrelevant to visual data, or if the complexity of the image is high, the output of the method would be quite unsatisfactory \cite{dummy:14}. In \cite{dummy:15}, low-light enhancement model using convolutional neural network and Retinex theory was proposed. It showed an equivalence between multi-scale Retinex and feedforward convolutional neural network using Gaussian kernels. However, because of the limited receptive field in their model, very smooth regions such as clear sky are sometimes attacked by the halo effect. Jeremias sulam, formulated trainlets, to construct large adaptable atoms using various datasets of facial images using dictionary learning algorithm. Because of the computational constraints, this method was applied to tiny regions of the image and not on the entire image. As a result, this approach did not give satisfactory results on large regions in images \cite{dummy:17}. 
Raymond and Chen \cite{dummy:19} proposed another picture completion technique that can be utilized to fix any state of gaps. In any case, such training depends on the data used in training. In the meantime, the processing of surface and structure was not sufficiently impeccable. Kai Zhang and Yunjin Chen proposed a picture denoising approach in which they built feed-forward denoising convolutional neural systems using residual learning and batch normalization. However, this methodology was unfit to recover missing regions, and it just denoised the picture. Likewise, it was unfit to refine pictures with genuine complex commotion and other general picture restoration tasks \cite{dummy:20}. Zhao, Liu, and Hiang proposed deep neural networks to inpaint and de-noise the corrupted image. However, in this method, the handling of structure and texture was not satisfactory \cite{dummy:21}.

\section{THE PROPOSED MODEL}
The methodology could be separated into three different steps. In the first step, data-preprocessing on CelebA-hq dataset \cite{dummy:10} is done to run and test the developed model. In the second step, a Wasserstein GAN based model to complete the missing pixels in the image is developed. The image completion GAN gives a complete image with a blurry filled area. The generator of the GAN generates real looking images, but in the process of generation, the noise gets unavoidably added. So, in the third step, the output of the generator is passed through the enhancement network to make the filled area clear and to refine the completed image further.

\subsection{DATA PREPROCESSING}
The following data preprocessing steps were followed:
\begin{itemize}
\item The dataset is splited into 15000 training images and around 1000 testing images.
\item Each face image in the dataset is resized to 64* 64* 3 pixels to train the Wasserstein GAN model.
\item Masking- A binary mask is used with values 0 or 1. 0 corresponds to the corrupted region while 1 corresponds to the uncorrupted region in the image. This binary mask is applied to all images  to make them corrupted which will serve as input of the training process.
\item The enhancement network is trained using 2000 image pairs containing blurry images and its corresponding clean images.  
\end{itemize} 

\subsection{WASSERSTEIN GAN}
The concept of GANs was put forward by Goodfellow\cite{dummy:11} which trains two networks simultaneously: the generator network G to learn the distribution of training data and the critic network C which distinguishes the generated samples from the original samples as in Figure 1. The generative network is trained to generate patches to complete the missing regions in the image. Meanwhile, it is difficult for the critic network to classify the output of the generator from the original dataset sample. The GAN architecture used is Wasserstein GAN, which uses Wasserstein distance as to train the generator so that it can capture training data distribution and generate images similar to those in the training data.  

\begin{figure}[h!]
	\begin{center}
		\includegraphics[scale=0.25]{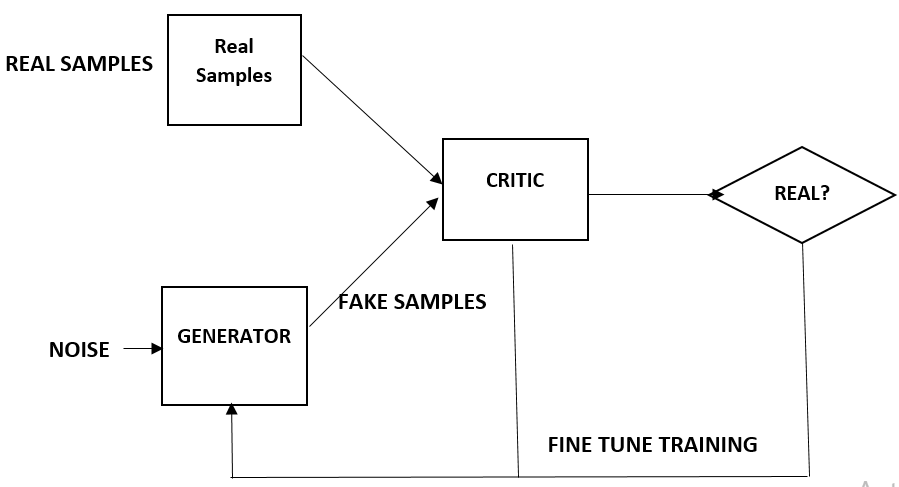}
		\caption{The Framework for Wasserstein GAN}
	\end{center}
\end{figure}

\subsection{WASSERSTEIN DISTANCE AS LOSS FUNCTION FOR THE TRAINING OF GENERATOR}
Wasserstein distance is a measure of the distance between two probability distributions.
For the generated data distribution $p_{g}$ and the real data distribution $p_{r}$, it can be mathematically defined as the cost for the cheapest plan from $p_{g}$ to $p_{r}$. It is also called {\bf Critic loss} or {\bf Wasserstein distance}. 
\noindent The Wasserstein distance loss function L to train the generator can be mathematically represented as\cite{dummy:16}:
\begin{equation}
L=\displaystyle\mathop{\mathbb{E}}_{\overset{\sim}{x}\sim P_{g}}[C(\overset{\sim}{x})]-\displaystyle\mathop{\mathbb{E}}_{x\sim P_{r}}[C(x)]    
\end{equation}
Here, the first term represents the expectation of the distribution generated by the generator, and the second term represents the expectation of the real training data distribution. By minimizing the difference between the two, the generator learns to generate samples having probability distribution similar to training data distribution.
\noindent Now, to make the learning faster and make model convergence faster gradient penalty term is added to our loss function. So, the overall loss function L of Wasserstein GAN becomes:
\begin{equation}
L=\displaystyle\mathop{\mathbb{E}}_{\overset{\sim}{x}\sim P_{g}}[C(\overset{\sim}{x})]-\displaystyle\mathop{\mathbb{E}}_{x\sim P_{r}}[C(x)]+gradient\,penalty
\end{equation}
where, gradient\,penalty will be given by
\begin{equation}
gradient\,penalty=\lambda\displaystyle\mathop{\mathbb{E}}_{\hat{x}\sim P_{g}}[(||\bigtriangledown_{\hat{x}} C(\hat{x})||_{2}-1)^2]   
\end{equation}
here $\lambda$ is the  gradient penalty coefficient.
\subsection{IMAGE COMPLETION WITH WASSERSTEIN GAN}
After training the generator to generate samples which look real, the next aim is to ensure that the missing region generated has a similar context to the non-missing region so that sensible looking completed images as output can be obtained. For this, the following is done:\\

\noindent A binary mask with values 0 or 1 is used. 0 corresponds to the corrupted region while 1 corresponds to the uncorrupted region in the image. Let y represents the uncorrupted image. $M\odot y$ gives the uncorrupted part of the image. Let G($z^{'}$) be some image generated by the generator which suitably completes the missing region in the image. (1-M) $\odot$ G($z^{'}$) represents the completed region which when added to the uncorrupted region gives the reconstructed image \cite{dummy:18} as output:

\begin{equation}
     x_{reconstructed} = M\odot y+(1-M)\odot G(z^{'})  
\end{equation}

\noindent To find $z^{'}$ that suitably completes the image following loss functions are defined \cite{dummy:18}:\\

\noindent {\bf Contextual Loss:} To ensure both generated and the input image have same context, ensure that the uncorrupted pixel  in original image y are same as the pixels in the generated image G(z) at a particular location. For this, pixel wise difference between the uncorrupted part of the two images is taken and then this difference is minimized.
\begin{equation}
     L_{contextual}(z) = ||M\odot G(z) - M\odot y ||_1 
\end{equation}
where $||x||_1$ represents $l_1$ norm of some vector x. \\
\noindent {\bf Perceptual Loss:} It  ensures that the output image looks real. For this, the following perceptual loss:
\begin{equation}
    L_{perceptual}(z) = log(1-C(G(z)))
\end{equation}
{\bf Total loss}:
It is a sum of perceptual and contextual loss and is denoted by L(z):
\begin{equation}
    L(z) = L_{contextual}(z) + Q L_{perceptual}(z)
\end{equation}
\noindent Q is a hyper-parameter and we minimize this loss function to ensure completed image is contextally similar to input image.

\subsection{ENHANCEMENT NETWORK USING RESIDUAL LEARNING}

\begin{figure}[h!]
	\begin{center}
		\includegraphics[scale=0.25]{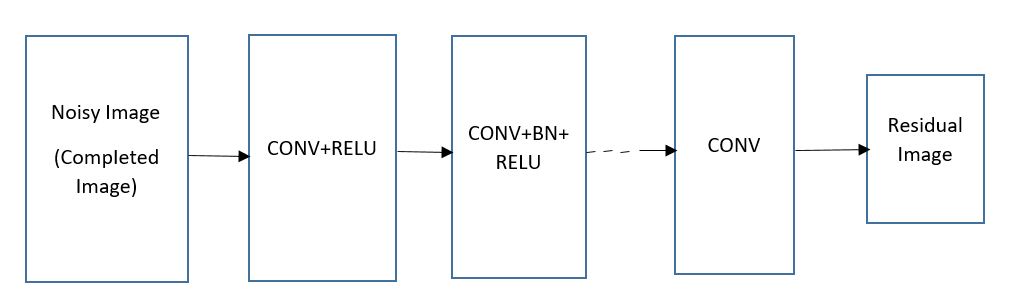}
		\caption{Enhancement network to refine completed images}
	\end{center}
\end{figure}
In enhancement network to refine completed images, the residual learning approach is used. The input to the network is blurry image y = x + v, here x is the clear image,v represents the blur added. The residual network is trained to grasp the mapping R(y)\( \approx \)v , to get the clear image x as x = y- R(y). 
Mathematically, the average mean square error among the output residual image by the model and the actual residual images is used as error function for getting the parameter \( \Theta \) to train the enhancement network. 
\begin{equation}
L(\theta)=\frac{1}{2N}\sum_{i=1}^{N}||R(y_{i};\theta)-(y_{i}-x_{i})||^2 
\end{equation}
Here, L is the training error of the enhancement network and N are total training images. \\

\noindent Enhancement network consists of following layers as shown in Figure 2: 
(i) Conv+ReLU: It creates feature maps, and ReLU adds the non-linearity.
(ii) Conv+BN+ReLU: This layers contains added batch normalization between Conv and ReLU.
(iii) Conv: It is used to get the output residual image.

\section{RESULTS AND DISCUSSION}

The following plot was obtained by training the enhancement network on 2000 celeba-hq image pairs of clean and its corresponding blurr images.

\begin{figure}[h!]
	\begin{center}
		\includegraphics[scale=0.55]{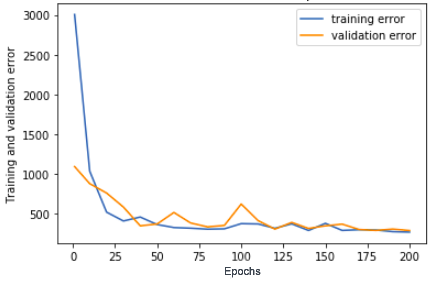}
		\caption{Enhancement network training plot to refine completed images}
	\end{center}
\end{figure}

In Figure 3, as the training proceeds, the average mean square error among the output residual image by the model and the actual residual images decreases. As a result, according to Equation (8), the training error decreases. Finally, around 200th epoch, the enhancement network is sufficiently trained ,which is evident as the training error becomes constant at a particular value, and there is no further decrease.\\

The following Wasserstein distance plot was obtained while training Wasserstein GAN on around 15000 Celeba-hq images for 10000 epochs and batch size of 128.

\begin{figure}[h!]
	\begin{center}
		\includegraphics[scale=0.55]{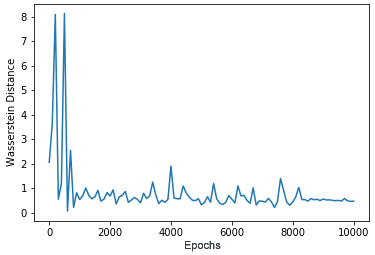}
		\caption{Wasserstein distance plot for training the generator}
	\end{center}
\end{figure}

In Figure 4, from Equation (2) it can be seen that in the initial stages of learning the expectation of the distribution generated by the generator is different from the expectation of the distribution of real data and hence the difference between the two is higher resulting in higher Wasserstein distance values. However, as the learning proceeds generator learns the distribution of the real data and then generates samples having a similar distribution with the real data, and hence the difference in their expectation decreases resulting in lower Wasserstein distance values. Now, around 10000 epochs the generator has sufficiently learned, and hence the Wasserstein distance values do not decrease further and becomes constant around a particular lower value.\\

The following contextual, perceptual and total loss plots were obtained while training the Wasserstein GAN for image completion for 1250 epochs on 15000 Celeba-hq images.

\begin{figure}[h!]
	\begin{center}
		\includegraphics[scale=0.55]{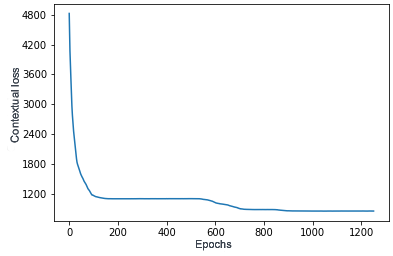}
		\caption{Contextual loss plot for image completion}
	\end{center}
\end{figure}

In Figure 5, in the initial stage, the context in the uncorrupted region of the generated samples and the original samples are different, so from Equation (5), it can be seen that the resulting contextual loss is higher. As the training moves further using Adam's optimizer (z) gets trained, and hence, there is a significant decrease in the contextual loss values. Around 1200th epoch, the context in the uncorrupted region of the generated samples and the original samples becomes quite familiar, and hence the contextual loss becomes constant around a particular value.

\begin{figure}[h!]
	\begin{center}
		\includegraphics[scale=0.55]{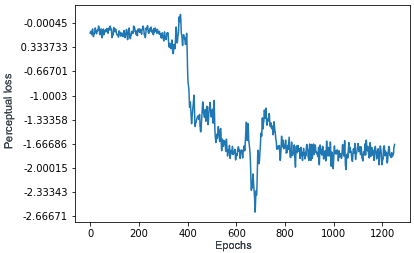}
		\caption{Perceptual loss plot for image completion}
	\end{center}
\end{figure}

In Figure 6, initially the distribution of generated images and real images is different, so the critic is able to distinguish the generated samples from the real ones and hence the value of C(G(z)) is close to 0 and as a result 1-C(G(z)) becomes close to 1 as a result from Equation (6) the loss is higher. However, as learning proceeds, the generator generates real looking samples as a result
C(G(z)) becomes close to 1 and 1-C(G(z)) becomes close to 0, resulting in lower perceptual loss values from Equation (6). 

\begin{figure}[hbt!]
	\begin{center}
		\includegraphics[scale=0.55]{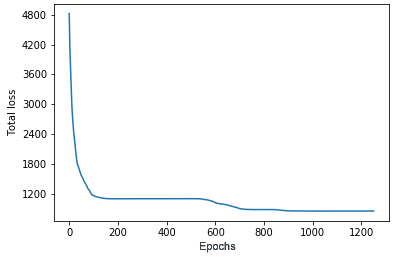}
		\caption{Total image completion loss plot}
	\end{center}
\end{figure}

The perceptual loss values are quite lower compared to contextual loss values, and as a result from Equation (7), the total image completion loss is almost equal to contextual loss, and as a result Figure 7 which is total image completion loss plot is almost similar to contextual loss for image completion plot Figure 5.

The following two evaluation metrics to evaluate the quality of the output images by the model:

\subsection{Peak Signal-to-Noise Ratio (PSNR):}

PSNR \cite{dummy:5} is measured in decibels (dB). The higher the PSNR, the better image has been completed to match the original image.
\begin{equation}
MSE=\frac{1}{mn}\sum_{i=0}^{m-1}\sum_{j=0}^{n-1}||f(i,j)-g(i,j)||^2 
\end{equation}
\begin{equation}
PSNR=20\log_{10}(\frac{MAX_{f}}{\sqrt{MSE}}) 
\end{equation}
\noindent Here,
f is the original image,
g represents completed image through the model,
m represents image pixel rows,
n represents image pixel columns, i and j represents row and column index respectively. MAX$_{f}$  is a constant equal to 255.

\section{{ Structural Similarity Index (SSIM):}}

\noindent The Structural Similarity (SSIM) Index \cite{dummy:5} depends on computation of terms, namely the luminance, contrast and structural term.
\begin{equation}
SSIM(x,y)=[C(x,y)]^\alpha\times[I(x,y)]^\beta\times[S(x,y)]^\gamma
\end{equation}

\noindent where, C represents contrast, I represents luminance, S represents structural term, x represents original, y represents completed images. The parameters $\alpha > 0$, $\beta > 0 $, and $\gamma > 0$, are used to adjust the relative importance of the three
components.

The following PSNR and SSIM values through the proposed approach:

\begin{table}
\centering
\caption{Comparison of PSNR values}
\begin{tabular}{|p{1.5cm}|p{1.3cm}|p{1.3cm}|p{1.5cm}|}
	\hline
 	\textbf{Methods} & \textbf{CE\cite{dummy:12}}  &\textbf{PI \cite{dummy:3}} & \textbf{This work}\\
	\hline
	\textbf{PSNR(dB)} &	22.85  & 21.45  &  23.41\\
	\hline
\end{tabular}
\end{table}
\begin{table}
\centering
\caption{Comparison of SSIM values}
\begin{tabular}{|p{1.5cm}|p{1.3cm}|p{1.3cm}|p{1.5cm}|}
	\hline
 	\textbf{Methods} & \textbf{CE\cite{dummy:12}}  &\textbf{PI \cite{dummy:3}}& \textbf{This work}\\
	\hline
	\textbf{SSIM} &	0.872  &  0.851  &  0.9074\\
	\hline
\end{tabular}
\end{table}
\noindent It can be seen that the approach performs well CelebA-hq dataset compared to other proposed image completion techniques, which is evident from the above PSNR and SSIM values.\\

Some of the results obtained through the proposed image completion approach using Wasserstein GAN are shown below in Figure 8(a-e):

\begin{figure}[h!]
\hspace{-0.10in} Original \hspace{0.53in} Input \hspace{0.57in} Output \\

\centering
\subfloat{\includegraphics[width = 0.60 in]{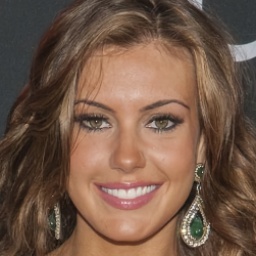}} \hspace{7.5 mm}
\setcounter{subfigure}{0}
\subfloat[]{\includegraphics[width = 0.60 in]{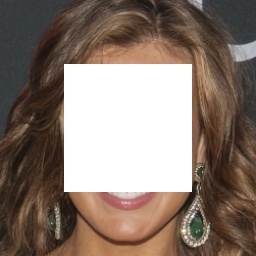}} \hspace{7.5 mm}
\subfloat{\includegraphics[width = 0.60 in]{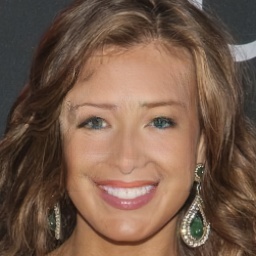}} \\ 
\subfloat{\includegraphics[width = 0.60 in]{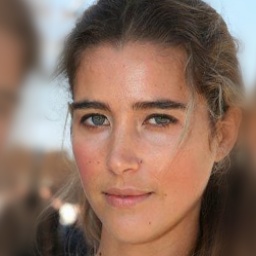}} \hspace{7.5 mm}
\setcounter{subfigure}{1}
\subfloat[]{\includegraphics[width = 0.60 in]{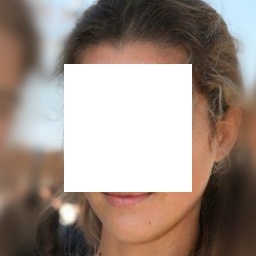}} \hspace{7.5 mm}
\subfloat{\includegraphics[width = 0.60 in]{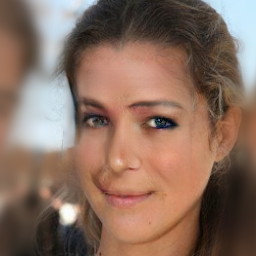}} \\ 
\subfloat{\includegraphics[width = 0.60 in]{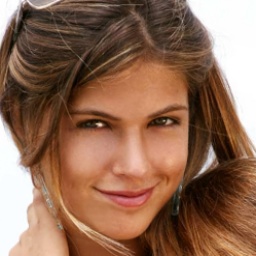}} \hspace{7.5 mm}
\setcounter{subfigure}{2}
\subfloat[]{\includegraphics[width = 0.60 in]{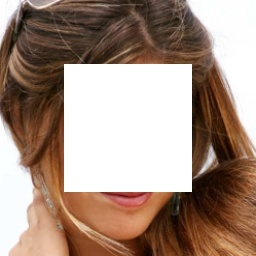}} \hspace{7.5 mm}
\subfloat{\includegraphics[width = 0.60 in]{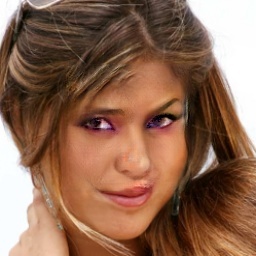}} \\ 
\subfloat{\includegraphics[width = 0.60 in]{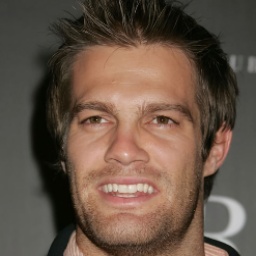}} \hspace{7.5 mm}
\setcounter{subfigure}{3}
\subfloat[]{\includegraphics[width = 0.60 in]{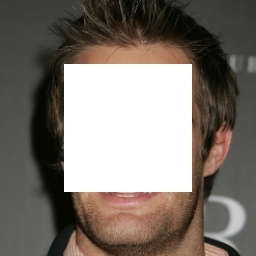}} \hspace{7.5 mm}
\subfloat{\includegraphics[width = 0.60 in]{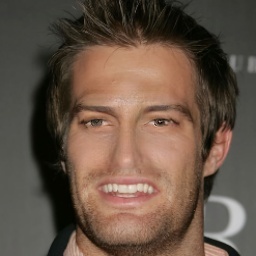}} \\ 
\subfloat{\includegraphics[width = 0.60 in]{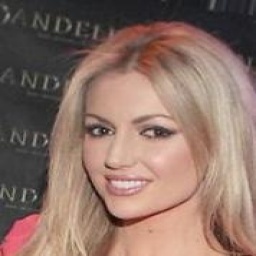}} \hspace{7.5 mm}
\setcounter{subfigure}{4}
\subfloat[]{\includegraphics[width = 0.60 in]{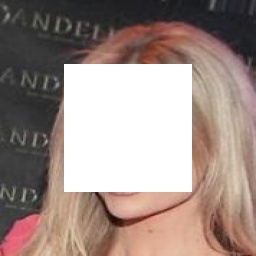}} \hspace{7.5 mm}
\subfloat{\includegraphics[width = 0.60 in]{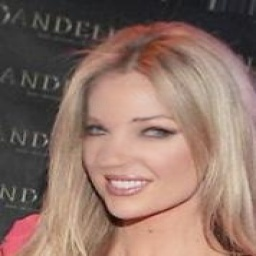}} \\ 
\caption{Experimental Results}
\label{some example}
\end{figure}

\section{CONCLUSION}
In this paper, the Wasserstein Generative Adversarial Network (WGAN) is first trained to generate the missing patches in the image and then passed the completed image given by the WGAN architecture is passed through an enhancement network to remove the blur and unwanted noise. By integrating image completion and enhancement task into a single process, the proposed approach provides better inpainting solutions by improving the Peak Signal to Noise ratio and Structural Similarity Index values by 2.45\% and 4\% respectively when compared to the recently used approaches. However, in this approach, overall training is highly depended on the data used for training. In future, work can be done towards optimizing the overall structure of the network and raise the network's capability to grasp minute details of the image further and to improve the image completion model.



\end{document}